\documentclass[12pt]{article}
\usepackage{amsfonts}
\usepackage{graphicx}
\usepackage{amsmath}
\usepackage{float}
\newcommand {\be}{\begin{equation}}
 \newcommand {\ee}{\end{equation}}
 \newcommand {\bea}{\begin{array}}
 
 \newcommand {\eea}{\end{array}}

\textwidth=6.5in \hoffset=-.75in \voffset=-1in \numberwithin{equation}{section}
\numberwithin{figure}{section}

\begin{document}

\begin{titlepage}
\vspace{1cm} 
\begin{center}
{\Large \bf {Instability of charged massive scalar fields in bound states around Kerr-Sen black holes}}\\
\end{center}
\vspace{2cm}
\begin{center}
\renewcommand{\thefootnote}{\fnsymbol{footnote}}
Haryanto M. Siahaan{\footnote{haryanto.siahaan@gmail.com}}
\\
Physics Department, Parahyangan Catholic University,\\
Jalan Ciumbuleuit 94, Bandung 40141, Indonesia

\renewcommand{\thefootnote}{\arabic{footnote}}
\end{center}

\begin{abstract}
We show the instability of a charged massive scalar field in bound states around Kerr-Sen black holes. By matching the near and far region solutions of the radial part in the corresponding Klein-Gordon equation, one can show that the frequency of bound state scalar fields contains an imaginary component which gives rise to an amplification factor for the fields. Hence, the unstable modes for a charged and massive scalar perturbation in Kerr-Sen background can be shown. 
\end{abstract}
\end{titlepage}\onecolumn 
\bigskip 

\section{Introduction}
\label{sec:intro}

The scalar field propagation in a rotating black hole background has been a subject of active investigations in theoretical physics community \cite{Kokkotas:1999bd,Supperrads-book}. Due to the centrifugal force, an impinging scalar wave can be partially reflected by the rotating black hole and scattered to infinity, while part of the wave is absorbed by the black hole. Interestingly, it was found that the reflection coefficient of the scalar wave that is reflected by the potential barrier of a charged rotating black hole can be greater than unity if the oscillation frequency of scalar fields is in the range
\be \label{0.1}
0 < \omega < m \Omega_H + e\Phi_H .
\ee 
In equation (\ref{0.1}), which is known as the superradiance condition, $m$ is the angular momentum of scalar field, $\Omega_H$ and $\Phi_H$ are the angular momentum and electrostatic potential of black hole at the horizon respectively. This phenomenon is known as superradiance effect, where the amplitude of reflected wave is amplified by the rotation of black holes. 

Related to superradiance effect, Press and Teukolsky \cite{teu-bomb} introduced the ``black hole bomb'' where the amplified scalar wave grows its amplitude without bound. The authors of \cite{teu-bomb} proposed that this is possible if a special arrangement of mirrors is placed in some regions outside the black hole which allow the scalar field to bounce back and forth between the mirrors and black hole. In this mechanism, the scalar wave amplitude gets amplification factor repeatedly each time reflected by the potential barrier of black holes. In the case of exponential amplification factor, then the scalar field is unstable. In ref. \cite{Cardoso:bomb} it is shown that there is a minimum distance where the mirror is placed from black hole to allow the existence of unstable modes.

However, it is found that nature can provide such an ``artificial'' mirror if the impinged scalar field is massive. The radial part of associated Klein-Gordon equation would be a Schrodinger-like, whose solutions represent a spectrum of bound states \cite{Damour,Furuhashi:2004jk} when a set of appropriate boundary condition is imposed. Naturally, the bound state's frequency is complex, where the imaginary part dictates the growing (decaying) in field's amplitude with time. As an example, in Ref. \cite{Furuhashi:2004jk} Furuhashi and Nambu showed that the bound state of a charged massive scalar field in the background of Kerr-Newman black hole is unstable. The authors managed to obtain the growth rate of scalar field's amplification analytically and numerically. 

It is known that a rotating and charged black hole solution is not an exclusive property in the Einstein-Maxwell theory. For example, in the low energy limit of heterotic string theory there is also a charged and rotating solution known as Kerr-Sen metric. It was Sen\cite{sen} who first derived this solution, by transforming the Kerr metric which also solves the equation of motion for $g_{\mu\nu}$ derived from the effective action in the low energy limit heterotic string theory. However, the rotating and electrically charged black hole solution in this theory has some distinguishable properties compared to the Kerr-Newman solution\cite{KS-KN}. For example, the authors of Ref. \cite{Hioki:2008zw} investigated the capture and scattering of photons by the Kerr-Sen and Kerr-Newman black holes, and also presented several examples which inform the characteristic differences of the capture region due to their spacetime structures. Also, the authors of Ref. \cite{Koga:1995bs} studied the evaporation of these black holes, and found their emission rates to be different. In addition to these examples, the hidden conformal symmetries of Kerr-Sen black holes is studied in Ref. \cite{KS-KN} where the authors found that one cannot have the twofold hidden conformal symmetries which Kerr-Newman black holes possesses\cite{Chen:2010ywa}. 

By exploring more on the physical aspects of Kerr-Sen black holes\cite{KS2}, one could gather more information that could be used to distinguish a charged and rotating black hole in string and Einstein-Maxwell theories. Hopefully this information can be challenged against astronomical observation and finally one can confirm the observed black hole belongs to Einstein-Maxwell theory or string theory. Specifically, a number of works in literature which explore the instability of massive scalar fields in Kerr-Newman background\cite{KN2,ranliKN} and also in static charged black holes in string theory\cite{ran-li,ran-li2,ran-li3} motivate us to perform such analysis for Kerr-Sen black holes. As one would expect, we find that the bound state of a charged massive scalar field in the background of Kerr-Sen black holes is also unstable, as in the Kerr-Newman case. To indicate the corresponding unstable modes, we adopt the analytical method by Furuhashi and Nambu\cite{Furuhashi:2004jk}. The two asymptotic radial solutions of scalar fields in Kerr-Sen background are matched and then the complex frequency of the bound states can be found. The results in this work support the finding that a (charged) massive scalar field is unstable in the background of a (charged) rotating black hole\cite{Damour,Supperrads-book}.

The paper is organized as follows. In section \ref{sec:KerrSen} we review the Kerr-Sen black holes and the corresponding minimally coupled curved spacetime scalar field equation. Subsequently, in section \ref{sec:sup}, we obtain the range of frequencies for a charged massive scalar field associated to the occurrence of superradiance effect. In section \ref{sec:near-region} we derive the radial solution in the near region, defined by $\omega(r-r_+)\ll 1$, and then in section \ref{sec:far-region} we solve the radial equation in far region, i.e. $r-r_+\gg M$. The solution in the matching region, $M\omega \ll\omega (r-r_+)\ll 1$, is given in section \ref{sec:matching}, from which the unstable modes can be shown in the succeeding section. In section \ref{sec:unstable} we also discuss on the instability of the scalar field's bound state when the limit $a\to 0$ is taken. As we know, Kerr-Sen solution reduces to the Gibbons-Maeda-Garfinkle-Horowitz-Strominger (GMGHS) metric when the rotation is turned off. In the last section, we summarize the paper and discuss some possible future works. Throughout this paper we use the units where $G = \hbar  = c = 1$.

\section{Massive Charged Scalar Field in Kerr-Sen Background}\label{sec:KerrSen}
In Ref. \cite{sen}, Sen derived a four dimensional solution that describes a rotating and electrically charged massive body in the low energy heterotic string field theory. The effective action in this theory reads
\be \label{action-low-het}
S = \int {d^4 x\sqrt {\left| g \right|} } \left( {R - \frac{1}{8}F^2  + \left( {\partial \tilde \Phi } \right)^2  - \frac{1}{{12}}H^2 } \right)\exp \tilde \Phi .
\ee
In the action above, $F^2$ is the squared of field-strength tensor $F_{\mu \nu }  = \partial _\mu  A_\nu   - \partial _\nu  A_\mu  $, $\tilde{\Phi}$ is the dilaton, and $H^2$ is the squared of a third rank tensor field,
\be
H_{\kappa \mu \nu }  = \partial _\kappa  B_{\mu \nu }  + \partial _\nu  B_{\kappa \mu }  + \partial _\mu  B_{\nu \kappa }  - \frac{1}{4}\left( {A_\kappa  F_{\mu \nu }  + A_\nu  F_{\kappa \mu }  + A_\mu  F_{\nu \kappa } } \right),
\ee 
where $B_{\mu\nu}$ is a second rank antisymmetric tensor field. In the absence of all non-gravitational fields, i.e. $A_\mu$, $\tilde{\Phi}$, and $B_{\mu\nu}$, the theory described by (\ref{action-low-het}) reduces to the vacuum Einstein gravitational theory where the Kerr metric is known as a rotating solution. In Ref. \cite{sen}, Sen made use of a transformation to the Kerr metric which leads to a new line element, namely the Kerr-Sen spacetime. 

In Boyer-Lindquist coordinates $(t,r,\theta,\phi)$, Kerr-Sen metric can be written as
\begin{eqnarray}
ds^2&=&-(1-\frac{2Mr}{\rho^2})dt^2+\rho^2(\frac{dr^2}{\Delta}+d\theta^2)-\frac{4Mra}{\rho^2}\sin^2\theta dtd\phi\nonumber\\
&&+\left(\rho^2+a^2\sin^2\theta+\frac{2Mr a^2\sin^2\theta}{\rho^2}\right)\sin^2\theta d\phi^2\label{KSmetric}.
\end{eqnarray}  
In the equation above, $\rho^2=r(r+2b)+a^2\cos^2\theta$, $\Delta=r(r+2b)-2Mr+a^2$, and $b=Q^2/2M$. The rotational parameter $a$ is defined as a ratio between black hole's angular momentum $J$ and its mass $M$. The black hole's electric charge is denoted by $Q$. The solutions for non-gravitational fundamental fields are\cite{KS1}
\begin{eqnarray}
{\tilde\Phi}&=&-\frac{1}{2}\ln \frac{{r}({r}+2b)+a^2\cos^2\theta}{{r}^2+a^2\cos^2\theta},\label{dil}\\
A_{{t}}&=&\frac{-{r}Q}{\rho ^2},\label{A-t}\\
A_{{\phi}}&=&\frac{{r}Qa\sin^2\theta}{\rho ^2},\label{A-phi}\\
B_{{t}{\phi}}&=&\frac{b{r}a\sin^2\theta}{\rho ^2}.\label{antisym-tensor-field}
\end{eqnarray}

Equation (\ref{KSmetric}) contains a black hole solution with the outer and inner horizons are located at $r_+=M-b+\sqrt{(M-b)^2-a^2}$ and $r_-=M-b-\sqrt{(M-b)^2-a^2}$ respectively. The corresponding Hawking temperature, angular velocity, and electrostatic potential at the horizon are respectively given by
\begin{eqnarray}
T_H&=&\frac{r_+-r_-}{8\pi Mr_+} =\frac{\sqrt{(2M^2-Q^2)^2-4J^2}}{4\pi M(2M^2-Q^2+\sqrt{(2M^2-Q^2)^2-4J^2})},\label{TH} \\
\Omega_H&=& \frac{a}{2Mr_+} =\frac{J}{M(2M^2-Q^2+\sqrt{(2M^2-Q^2)^2-4J^2})},\label{Omeg}\\
\Phi_H&=&\frac{Q}{2M}\label{Electrostatic_H}.
\end{eqnarray}

Setting the parameter $b=0$ yields all non-gravitational fields (\ref{dil}) - (\ref{antisym-tensor-field}) vanish, and the line element (\ref{KSmetric}) reduces to the Kerr metric. Moreover, turning off the rotational parameter $a$ transforms the Kerr-Sen spacetime (\ref{KSmetric}) to the GMGHS (Gibbons-Maeda-Garfinkle-Horowitz-Strominger) solution\cite{GMGHS}, which describes a spacetime outside of an electrically charged mass in string theory. 

Now we consider a massive and electrically charged scalar field $\Phi$ outside of a Kerr-Sen black hole. The corresponding equation for $\Phi$ can be read as  
\begin{equation}
    \left( {\nabla _\mu   - ieA_\mu  } \right)\left( {\nabla ^\mu   - ieA^\mu  } \right)\Phi  = \mu^2 \Phi,\label{KGeqtn}
\end{equation}
where $e$ and $\mu$ are the electric charge and mass of $\Phi$ respectively, and vector field $A_\mu$ is given by (\ref{A-t}) and (\ref{A-phi}). For the scalar field we use the ansatz
\begin{equation}
\Phi \left( {r,t,\theta ,\phi } \right) = e^{- i\omega t+im\phi }S\left( \theta  \right)R\left( r \right).\label{Phi-separation}
\end{equation}
Plugging eq. (\ref{Phi-separation}) into eq. (\ref{KGeqtn}) gives us an equation that furthermore can be separated into two parts, namely the angular part
\be\label{eq-ang}
\partial _\theta  \left( {\sin \theta \partial _\theta  S\left( \theta  \right)} \right) + \sin \theta \left( {\mu ^2 a^2 \sin ^2 \theta  - \frac{{m^2 }}{{\sin ^2 \theta }} + a^2\omega^2 \cos^2\theta + \lambda } \right)S\left( \theta  \right) = 0\,,
\ee
and the radial one
\be\label{eq-rad}
\partial _r \left( {\Delta \partial _r R\left( r \right)} \right) + \left( {\frac{{\left( {\omega \left( {\Delta  + 2Mr} \right) - eQr - am} \right)^2 }}{\Delta } - \mu ^2 \left( {\Delta  + 2Mr} \right) - \lambda '} \right)R\left( r \right) = 0\,,
\ee 
where $\lambda' = \lambda -2am\omega$. In this paper we consider the scalar field whose mass is around its frequency, i.e. $\mu \approx \omega$, hence eq. (\ref{eq-ang}) has a spherical harmonics equation form with $\lambda = l(l+1)$.

\section{Superradiant bound}\label{sec:sup}

To show the range of frequencies of bound states scalar fields that gives rise to the superradiance effect, we need to study the asymptotic behavior of radial solution $R(r)$ at $r\to r_+$ and $r\to \infty$. To do so, it is convenient to introduce a new radial function
\be\label{new-radf} 
R\left( r \right) = \frac{{{Y}\left( {r_* } \right)}}{{\left( {\Delta  + 2Mr} \right)^{1/2} }},
\ee
together with a transformation
\be\label{drsdr}
\frac{{dr_* }}{{dr}} = \frac{{\left( {\Delta  + 2Mr} \right)}}{\Delta }.
\ee 
Then, by using (\ref{new-radf}) and (\ref{drsdr}), the radial equation (\ref{eq-rad}) can be rewritten as
\be \label{eq-rad-transformed}
\frac{{d^2 {Y}}}{{dr_*^2 }} - \left( {G^2  + \frac{{dG}}{{dr_* }}  + \frac{{\mu ^2 \Delta }}{{\left( {\Delta  + 2Mr} \right)}} - \frac{{K^2 -\Delta \lambda '}}{{\left( {\Delta  + 2Mr} \right)^2 }}} \right){Y} = 0
\ee
where $G = (r+b) \Delta/ (\Delta + 2Mr)^{2}$ and $K = \omega(\Delta + 2Mr)-eQr-am$. 

Taking $r\to \infty$ yields eq. (\ref{eq-rad-transformed}) reduces to
\be \label{eqrad-far}
\frac{{d^2 {Y}}}{{dr_*^2 }} + \left( \omega^2  - \mu^2  \right) {Y} = 0.
\ee
The solution to (\ref{eqrad-far}) is
\be \label{Y-far}
{Y} = e^{-ir_*\sqrt {\omega ^2  - \mu ^2 }}+{\cal R}e^{ir_*\sqrt {\omega ^2  - \mu ^2 }},
\ee  
where we have introduced ${\cal R}$ as the reflectance coefficient\cite{konoplya-rmp}. The solution (\ref{Y-far}) satisfies the ``bound states'' boundary condition where at $r\to \infty$ the scalar wave function $\Phi \to 0$. 

Now let us discuss the near outer horizon case, i.e. $r\to r_+$. In this consideration, $\Delta\to 0$, and eq. (\ref{eq-rad-transformed}) accordingly can be read as
\be \label{eq-rad-nearhor}
\frac{{d^2 Y}}{{dr_*^2 }} + \left( {\omega  - e\Phi _H  - m\Omega _H } \right)^2 Y = 0\,.
\ee 
By employing the ingoing boundary condition at $r\to r_+$, we have
\be\label{Y-near}
Y = {\cal T} e^{ - i\left( {\omega  - e\Phi _H  - m\Omega _H } \right)r_* }
\ee 
that solves eq. (\ref{eq-rad-nearhor}), where ${\cal T}$ is the transmittance coefficient\cite{konoplya-rmp}. It is clear that asymptotic solutions (\ref{Y-far}) and (\ref{Y-near}) represent the bound states of a scalar field in a black hole geometry, decaying at $r\to \infty$ and ingoing near the horizon.

In general, the radial equations (\ref{eqrad-far}) and (\ref{eq-rad-nearhor}) are just the Schrodinger-like equation
\be 
\frac{{d^2 Y}}{{dr_*^2 }} + \left( {\omega ^2  - V_{eff} } \right)Y = 0,
\ee 
where $V_{eff}$ is real. Therefore, the associated Wronskian
\be\label{Wrons}
{\cal W}\left( {Y,Y^* } \right) \equiv Y\frac{{dY^* }}{{dr_* }} - Y^* \frac{{dY}}{{dr_* }}
\ee 
obeys the continuity equation
\be \label{Wronsk-cond}
\frac{d{\cal W}(Y,Y^*)}{dr_*}=0.
\ee 
Furthermore, by employing the condition (\ref{Wronsk-cond}) to the solutions (\ref{Y-far}) and (\ref{Y-near}), we obtain a relation between the reflectance ${\cal R}$ and transmittance ${\cal T}$ as
\be\label{superrads}
\left| {\cal R} \right|^2  = 1 + \left(\frac{e\Phi _H  + m\Omega _H -\omega}{\omega}\right)\left| {\cal T} \right|^2 .
\ee 

Consequently, from eq. (\ref{superrads}) we understand that in the range of frequency
\be\label{superrads-freq}
0 < \omega  < e\Phi _H  + m\Omega _H ,
\ee
the scattered scalar field in bound state from black hole brings some amount of energy which is larger compared to the incident wave. This result is in agreement with a conclusion that can be drawn from the entropy increasing law for Kerr-Sen black hole\cite{KS-KN},
\be \label{mass-law}
\delta M - \Omega_H \delta J - \Phi_H \delta Q \geq 0 \,.
\ee 
In the case of interaction between a massive charged scalar field and Kerr-Sen black hole, one can show a relation\cite{Novikov}
\be \label{delJdelM}
\delta J = \frac{m\delta M}{\omega} ~~~~,~~~~\delta Q = \frac{e\delta M}{\omega}\,.
\ee
By plugging the last equation into (\ref{mass-law}), it is understood that an impinged scalar wave on black hole whose frequency satisfy (\ref{superrads-freq}) may acquire some amount of energy from the black hole. This phenomenon is known as the superradiance effect which is noticed first in the studies of fields perturbation in Kerr background\cite{superrad}. 

As we expect, the condition for bound state scalar wave frequencies (\ref{superrads-freq}) is exactly the same with the one found in Kerr-Newman case\cite{Furuhashi:2004jk,ranliKN}. It is not surprising since Kerr-Newman and Kerr-Sen black holes are quite similar in several physical aspects. They both describe the spacetime outside of an electrically charge and rotating collapsing object. One is in the context of Einstein-Maxwell theory, and the other is in the low energy heterotic string field theory. In the next sections, we will analyze the radial equation (\ref{eq-rad}) in the near, far, and matching regions, and also get the corresponding approximate solutions.

\section{Near region analysis}\label{sec:near-region}

The near region of a black hole is space outside of the outer horizon with radial position defined by $r-r_+\ll \omega^{-1}$. Accordingly, this condition also means that $\omega r  \ll 1$ since $r_+ \sim M$ and from the beginning we have confined our discussions to the case of low frequency scalar fields, $M\omega\ll 1$. Together with the condition $\mu \approx \omega$, eq. (\ref{eq-rad}) can be expressed in the near region as
\be\label{eq-rad-near-r}
\Delta \partial _r \left( {\Delta \partial _r R} \right) + \left( {\left( {\omega \left( {r_ +  \left( {r_ +   + 2b} \right) + a^2 } \right) - eQr_ +   - am} \right)^2  - l\left( {l + 1} \right)\Delta } \right)R = 0.
\ee 
Since $a\omega \ll 1$, we can approximate $\lambda ' \simeq l(l+1)$ as appeared in the last equation. 

Furthermore, to simplify the radial equation in the near region, we follow Ref. \cite{ranliKN} to introduce a new radial coordinate,
\be\label{z-def}
z = \frac{\Delta }{{\left( {r - r_ -  } \right)^2 }}.
\ee 
In this new coordinate, eq. (\ref{eq-rad-near-r}) transforms to 
\be\label{eq-rad-near-z}
z\partial _z \left( {z\partial _z R} \right) + \left( {\varpi ^2  - l\left( {l + 1} \right)\frac{z}{{\left( {1 - z} \right)^2 }}} \right)R = 0,
\ee 
where
\be\label{omega-hat}
\varpi  = \frac{{\omega \left( {r_ +  \left( {r_ +   + 2b} \right) + a^2 } \right) - eQr_ +   - am}}{{r_ +   - r_ -  }}.
\ee 
Interestingly, eq. (\ref{eq-rad-near-z}) can be transformed to a type of hypergeometric equation by introducing a new function
\be \label{ZtoR}
Z\left( z \right) = z^{ - i\varpi } \left( {1 - z} \right)^{ - l - 1} R\left( z \right).
\ee 
In terms of $Z\left( z \right)$, eq. (\ref{eq-rad-near-z}) reads
\be\label{eq-near-hor-funcZ}
z\left( {1 - z} \right)\partial _z^2 Z + \left[ {1 + 2i\varpi  - z\left( {2l + 3 + 2i\varpi } \right)} \right]\partial _z Z - \left( {l + 1} \right)\left( {l + 1 + 2i\varpi } \right)Z = 0.
\ee 

In the neighborhood $z = 0$, which is compatible with our near region analysis, the last equation can be solved by using the hypergeometric function
\[
Z(z) = {\cal A}\,_2 F_1 \left( {l + 1 + 2i\varpi ,l + 1,1 + 2i\varpi ;z} \right)
\]
\be
 ~~~~~~~~~~~~~~~~~~~~~~~~+ {\cal B}z^{ - 2i\varpi }\, _2 F_1 \left( {l + 1,l + 1 - 2i\varpi ,1 - 2i\varpi ;z} \right),
\ee 
where ${\cal A}$ and ${\cal B}$ are some constants. Plugging back the last equation to (\ref{ZtoR}) gives 
\be \label{R-sol-near}
R\left( z \right)= {\cal B}\left( {1 - z} \right)^{l + 1} z^{ - i\varpi } \,{_2 F_1} \left( {l + 1,l + 1 - 2i\varpi ,1 - 2i\varpi ;z} \right),
\ee
where we have set ${\cal A} = 0$ to get an agreement with the asymptotic form of scalar wave function (\ref{Y-near}).

In the next section, we will discuss the far region version of eq. (\ref{eq-rad}) and obtain the related solution. To get an approximate solution which is valid in both near and far regions, we can match the large $r$ limit of the near region solution to the small $r$ limit of the far region solution. This method is known as the matching technique, and it provides us a solution that is valid in both regions. After obtaining the matching solution, the instability of scalar fields can be shown by getting the non-zero imaginary part in scalar's frequencies. 

Therefore, now we need to get the large $r$ limit of (\ref{R-sol-near}). It can be done by taking $z\to 1$, since in the limit of large $r$ we have $(r-r_+) \simeq (r-r_-)$. Before we show the near region radial solution with large $r$, first we need to apply the $z\to 1-z$ transformation to the ingoing part of eq. (\ref{R-sol-near})\cite{Bailey}
\[
_2 F_1 \left( {l + 1,l + 1 - 2i\varpi ,1 - 2i\varpi ;z} \right) = \frac{\Gamma \left( {1 - 2i\varpi } \right)}{\left( {1 - z} \right)^{ 2l + 1}} \left[ {\frac{{\Gamma \left( { - 2l - 1} \right)\left( {1 - z} \right)^{2l + 1} }}{{\Gamma \left( { - l - 2i\varpi } \right)\Gamma \left( { - l} \right)}}} \right.
\]
\be\label{F1-z}
~~~~~~~~~~~~~~~~~~~~~~~~~~~~~~~~~~~~~~~~~~~~ \left. { + \frac{{\Gamma \left( {2l + 1} \right)}}{{\Gamma \left( {l + 1 - 2i\varpi } \right)\Gamma \left( {l + 1} \right)}}} \right].
\ee 
Accordingly, the reading of near region ingoing radial solution in the large $r$ condition becomes
\be\label{R-sol-near-largefinal}
R \sim {\cal A}\Gamma \left( {1 - 2i\varpi } \right)\left[ {\frac{{\Gamma \left( { - 2l - 1} \right)\left( {r_ +   - r_ -  } \right)^{l + 1} }}{{\Gamma \left( { - l - 2i\varpi } \right)\Gamma \left( { - l} \right)r^{l + 1} }} + \frac{{\Gamma \left( {2l + 1} \right)\left( {r_ +   - r_ -  } \right)^{ - l} r^l }}{{\Gamma \left( {l + 1} \right)\Gamma \left( {l + 1 - 2i\varpi } \right)}}} \right].
\ee

\section{Far region analysis}
\label{sec:far-region}

Now we look for a solution to the radial equation (\ref{eq-rad}) in the far region approximation, i.e. $r-r_+\gg M$. In such consideration, eq. (\ref{eq-rad-transformed}) reads 
\be\label{eq-rad-far}
\frac{{d^2 Y}}{{dr^2 }} + \left[ {k^2  + \frac{{2\left( {M\left( {2\omega ^2  - \mu ^2 } \right) - \omega eQ} \right)}}{r} - \frac{{l\left( {l + 1} \right)}+\varepsilon^2}{{r^2 }}} \right]Y = 0,
\ee 
where $k^2=\omega^2-\mu^2$, $M\omega\sim \varepsilon \ll 1$, and $eQ\sim \varepsilon\ll 1$. We keep the $\varepsilon^2$ term in the equation above in order to get an explicit $\varepsilon^2$ dependence of the solution. However, in the final result the ${\cal O}(\varepsilon^2)$ terms will be neglected. We will find that this way is useful in the next section, to guarantee that we do not encounter any singular valued Gamma function in our computation. It is not surprising that we find eq. (\ref{eq-rad-far}) has the form of far region radial equation in Kerr-Newman background for a charged massive scalar that appeared in Ref. \cite{Furuhashi:2004jk}. Again, clearly this is because both Kerr-Sen and Kerr-Newman black holes are examples of gravitational collapse solutions with rotation and electric charge. 

Hence, the method developed in Ref. \cite{Furuhashi:2004jk} to obtain the far region solution, as well as its matching technique, also applies to our case. However, we find that there is a freedom left in dealing with parameter $k$, i.e. transforming $k$ to $-k$ does not change eq. (\ref{eq-rad-far}), which furthermore affects another parameters which contain $k$.  A general solution that solves eq. (\ref{eq-rad-far}) is the combination of Whittaker functions\footnote{See Appendix \ref{app:Witt}.}
\be
Y = D_1 M_{\alpha ,\beta } \left( {2ikr} \right) + D_2 W_{\alpha ,\beta } \left( {2ikr} \right),
\ee 
where
\be \label{alp-bet}
\alpha  =  -i\frac{{\left( {M\left( {2\omega ^2  - \mu ^2 } \right) - \omega eQ} \right)}}{k},
\ee
and
\be 
\beta  = l + \frac{1}{2}+\frac{\varepsilon^2}{2l+1}.
\ee 

For the radial function $R(r)$ in (\ref{new-radf}) which can be read approximately as $r^{-1}Y$ in the case of large $r$, in order to satisfy the Dirichlet boundary condition at $r\to\infty$, the coefficient $D_1$ must vanish. Therefore, the solution satisfying the boundary condition reads
\[
R\sim e^{ - ikr} r^{ - 1} \left( {2ikr} \right)^{\beta  + 1/2} \left[ {\frac{{\Gamma \left( { - 2\beta } \right)M\left( {\beta  - \alpha  + 1/2;2\beta  + 1;r} \right)}}{{\Gamma \left( { - \beta  - \alpha  + 1/2} \right)}}} \right.
\]
\be\label{R-sol-far} 
\left. { + \frac{{\Gamma \left( {2\beta } \right)M\left( { - \beta  - \alpha  + 1/2; - 2\beta ;r} \right)r^{ - 2\beta } }}{{\Gamma \left( {\beta  - \alpha  + 1/2} \right)}}} \right].
\ee
Moreover, by using the identity $\Gamma \left( x \right)\Gamma \left( { - x} \right) =  - \pi \left( {x\sin \left( {\pi x} \right)} \right)^{ - 1} $ and the asymptotic form at $r\ll 1$ of the confluent hypergeometric function, $M(a;b;r)\sim 1$, the small $r$ approximation of eq. (\ref{R-sol-far}) is
\[
R\sim r^{ - 1} \left( {2ikr} \right)^{\beta  + 1/2} \frac{\pi }{{\sin \left( {2\beta \pi } \right)}}\left[ {\frac{1}{{\Gamma \left( {2\beta +1} \right)\Gamma \left( { - \beta  - \alpha  + 1/2} \right)}}} \right.
\]
\be\label{R-sol-far-nearR}
\left. { - \frac{{\left( {2ikr} \right)^{ - 2\beta } }}{{\Gamma \left( { 1 - 2\beta } \right)\Gamma \left( {\beta  - \alpha  + 1/2} \right)}}} \right].
\ee

\section{Solution in the matching region}\label{sec:matching}

In the matching region, defined by $M\omega \ll\omega (r-r_+)\ll 1$, the solutions (\ref{R-sol-near-largefinal}) and (\ref{R-sol-far-nearR}) overlap each other. Hence, by matching these two solutions, one can show that
\[
- \frac{{\Gamma \left( { - 2l - 1} \right)\left( {\left( {r_ +   - r_ -  } \right)\left( {2ik} \right)} \right)^{2l + 1} \Gamma \left( {l + 1 - 2i\varpi } \right)\Gamma \left( {l + 1} \right)}}{{\Gamma \left( { - l - 2i\varpi } \right)\Gamma \left( { - l} \right)\Gamma \left( {2l + 1} \right)}}
\]
\be \label{rhlh}
 = \frac{{\Gamma \left( {2\beta  + 1} \right)\Gamma \left( { - \beta  - \alpha  + 1/2} \right)}}{{\Gamma \left( {1 - 2\beta } \right)\Gamma \left( {\beta  - \alpha  + 1/2} \right)}}.
\ee
The right hand side of eq. (\ref{rhlh}) may vanish, provided that the argument of Gamma function in the denominator is some negative integers. Let us consider that it occurs in the case of $\alpha = \alpha_0$, where $\alpha_0 = n+\beta +1/2$ and $n$ is some nonzero positive integers. From eq. (\ref{alp-bet}), by considering that $\alpha_0$ depends on $\omega_0$ dan $k_0$, we have
\be
\alpha _0  =  - i\frac{{M\left( {2\omega _0^2  - \mu ^2 } \right) - eQ\omega _0 }}{{k_0 }} \simeq n + l + 1.
\ee 

Since $\alpha_0$ approximately must be some positive integers, and the condition $0 < \arg k < \pi $ which must be satisfied\footnote{Hence the solution $Y(r)$ in eq. (\ref{eq-rad-far}) obeys the boundary condition at $r\to\infty$.}, then from the last equation one understands that 
\be \label{eQmM}
eQ\omega _0  > M\left( {2\omega _0^2  - \mu ^2 } \right)\,.
\ee  
Then the associated $\omega_0$ and $k_0$ can be determined\cite{Furuhashi:2004jk} accordingly, approximately as
\be \label{omega0}
\omega _0  \simeq \mu \left( {1  -  \left( {\frac{{eQ - \mu M}}{{n + l + 1}}} \right)^2 } \right)^{\tfrac{1}{2}} \simeq \mu \left( {1  -  \frac{1}{2}\left( {\frac{{eQ - \mu M}}{{n + l + 1}}} \right)^2 } \right),
\ee
and
\be 
k_0  = \left(\omega_0^2-\mu^2\right)^{\tfrac{1}{2}} \simeq i\mu \left( {\frac{{eQ - \mu M}}{{n + l + 1}}} \right).
\ee
The last equation can be re-expressed as
\be
k_0  =  i\left| {k_0 } \right|\,,
\ee 
where
\be
\left| {k_0 } \right| = \mu \left({\frac{{eQ - \mu M}}{{n + l + 1}}}\right)\,.
\ee 
The right hand side of last equation is guaranteed to be positive by eq. (\ref{eQmM}).

Now we evaluate eq. (\ref{rhlh}) at $\alpha = \alpha_0 - \delta\alpha$, where $\delta\alpha$ is a small perturbation of $\alpha$. After a simple rearrangement, we have
\[
\left( { - 1} \right)^l \left( {\left( {r_ +   - r_ +  } \right)2k} \right)^{2l+1} 2\varpi \prod\limits_{j = 1}^l {\left( {j^2  + 4\varpi ^2 } \right)} 
\]
\be\label{far-near-coeff-match}
~~~~~~~~~~~~~~~~~ = 2l\left( {\frac{{2l!}}{{l!}}} \right)^2 \frac{{\Gamma \left( {2l + 2 + \tfrac{2\varepsilon ^2}{2l+1} } \right)\Gamma \left( {\delta\alpha  - n - 2l - 1-\tfrac{\varepsilon ^2}{2l+1}} \right)}}{{\Gamma \left( { - 2l - \tfrac{2\varepsilon ^2}{2l+1} } \right)\Gamma \left( {\delta\alpha  - n+\tfrac{\varepsilon ^2}{2l+1}} \right)}}
\ee 
Furthermore, some algebraic manipulations and the formula
\be 
\Gamma \left( {\varepsilon  - n} \right) \simeq \frac{{\left( { - 1} \right)^n }}{{n!}}\left( {\frac{1}{\varepsilon } + \Psi \left( {n + 1} \right)} \right)
\ee
which is valid for $\varepsilon \ll 1$ yield eq. (\ref{far-near-coeff-match}) can be read approximately as
\be \label{del-alpha}
\delta \alpha  \simeq \frac{{-i \left( {n + 2l + 1} \right)!}}{{n!\left( {2l!} \right)^2 }}\left( {\frac{{l!}}{{2l!}}} \right)^2 \frac{{\varpi_0 \left( {\left( {r_ +   - r_ -  } \right)2\left| {k_0 } \right|} \right)^{2l+1} }}{{l\left( {2l + 1} \right)}}\prod\limits_{j = 1}^l {\left( {j^2  + 4\varpi_0 ^2 } \right)}.
\ee
To get eq. (\ref{del-alpha}), we have neglected the ${\cal O}(\varepsilon^2)$ terms and use a new notation 
\be\label{omega-hat-0}
\varpi_0  = \frac{{\omega_0 \left( {r_ +  \left( {r_ +   + 2b} \right) + a^2 } \right) - eQr_ +   - am}}{{r_ +   - r_ -  }}.
\ee 
Also, since we consider $\alpha = \alpha_0 - \delta\alpha$ in the right hand side of eq. (\ref{far-near-coeff-match}), we should expressed $k$ as $k_0+\delta k$ and $\varpi$ as $\varpi_0 +\delta\varpi$ in the left hand side of the eq. (\ref{del-alpha}) accordingly. The quantities $\delta k$ and $\delta \varpi$ are understood as some small perturbations of $k$ and $\varpi$ respectively. However, the contribution of these small perturbation terms can be safely neglected. 

Now, having obtained $\omega_0$, $k_0$, and $\alpha_0$, inserting $\omega = \omega_0 +\delta\omega$ into (\ref{alp-bet}) gives us
\be\label{alpha-in-a0+dw}
\alpha  = \alpha _0  - \delta \omega \frac{{M\omega _0  - eQ}}{{ik_0^3 }}\mu ^2 .
\ee 
Consequently, it follows that
\be\label{del-omega1}
\delta \omega  \simeq \delta \alpha \frac{{\left( {M\mu  - eQ} \right)^2 }}{{\left( {l + n + 1} \right)^3 }}\mu \,,
\ee 
which can be read in more detail as
\be\label{del-omega2}
\delta \omega  \simeq \frac{{-i\mu {\left( {M\mu  - eQ} \right)^2 }\left( {n + 2l + 1} \right)!}}{{n!\left( {2l!} \right)^2 {\left( {l + n + 1} \right)^3 } }}\left( {\frac{{l!}}{{2l!}}} \right)^2 \frac{{\varpi_0 \left( {\left( {r_ +   - r_ -  } \right)2\left| {k_0 } \right|} \right)^{2l+1} }}{{l\left( {2l + 1} \right)}}\prod\limits_{j = 1}^l {\left( {j^2  + 4\varpi_0 ^2 } \right)}\,.
\ee 

The last equation deserves some comment. The range of frequencies which associate to the unstable modes of the bound state scalar perturbations is expressed in eq. (\ref{superrads-freq}). In this range, it is possible that $\varpi_0$ in eq. (\ref{omega-hat-0}) to be negative, which provides the sign of $\delta\omega$ in eq. (\ref{del-omega1}) to be positive. In the next section we will discuss how this positive and pure imaginary $\delta\omega$ is responsible in the amplification of scalar fields $\Phi$.

\section{The unstable modes}\label{sec:unstable}

In section \ref{sec:KerrSen}, we have considered the dependence of scalar wave with respect to time as $e^{-i\omega t}$. As we have shown in the previous section, the bound state frequency $\omega$ consists of the real and imaginary parts, namely $\omega_0$ and $\delta\omega$ respectively. From this fact, we realize that the scalar field could be either amplified or damped depending on the sign of $\delta\omega$ in eq. (\ref{del-omega2}). Suppose that we have $\delta\omega$ being negative and imaginary, then the scalar has a damping factor $e^{ - \left| {\delta \omega } \right| t} $. In contrary, if $\delta\omega$ is imaginary and positive, then we have $e^{\left| {\delta \omega } \right|t} $ which acts as the amplification factor of the scalar, i.e. the field's amplitude growth rate. In our present paper, the bound state scalar fields whose frequency\footnote{Recall that the conditions $M\mu\ll 1$ as well as $eQ\ll 1$ convince us that $\omega_0$ in eq. (\ref{omega0}) is real and positive.} $\omega_0$ is within the superradiant bound may get an amplification factor provided that $\varpi_0$ in eq. (\ref{del-alpha}) being negative. In fact, the negativity of $\varpi_0$ is somehow guaranteed by the frequency bound $0< \omega_0 <m\Omega_H + e\Phi_H$. At this point, we have seen how the superradiant instability for scalar fields in bound state can take place in Kerr-Sen background, as one would expect from such a rotating black holes. 

In a series of paper by Ran Li et al, Refs. \cite{ran-li,ran-li2,ran-li3}, the authors discuss the instability of charged and static black hole in string theory. They showed that the superradiance effect exists for GMGHS black holes, i.e. charged black holes in string theory analogous to the Reissner-Nordstrom black holes in Einstein-Maxwell theory. The metric describing GMGHS black holes can be obtained from (\ref{KSmetric}) after setting\footnote{Accordingly, the radial part of equation of motion for charged massive scalar field in this paper also reduces to that in Ref. \cite{ran-li}.} $a\to 0$ and shift the radial coordinate $r\to r - 2b$. Our result in eq. (\ref{del-omega2}) support the claim of paper \cite{ran-li} which shows the superradiant of scalar fields in GMGHS background. It can be seen by setting $a\to 0$ in eq. (\ref{del-omega1}) which gives us
\be \label{del-alpha-a0}
\delta \alpha'  \simeq \frac{{-i \left( {n + 2l + 1} \right)!}}{{n!\left( {2l!} \right)^2 }}\left( {\frac{{l!}}{{2l!}}} \right)^2 \frac{{\varpi'_0 \left( {2r'_ + \left| {k_0 } \right|} \right)^{2l+1} }}{{l\left( {2l + 1} \right)}}\prod\limits_{j = 1}^l {\left( {j^2  + 4{\varpi'}_0 ^2 } \right)},
\ee
where
\be 
r'_ +   = 2\left( {M - b} \right)\,\,\,{\rm and}\,\,\,\varpi '_0  = \frac{{2M\left( {M - b} \right)\omega _0  - eQM}}{{\left( {M - b} \right)}}.
\ee 

The superradiant condition for static and electrically charged black holes\footnote{It applies for RN and GMGHS black holes.}
\be 
0 < \omega_0  < e\Phi _H 
\ee 
yields $\varpi'$ could be real negative shows the possibility of amplification of charged scalar wave in the background of GMGHS black holes. It is also easy to see that results presented in this paper tell us that the superradiance effect may occur in the case of neutral massive scalar perturbation in Kerr-Sen background, if its frequency is within the range
\be 
0 < \omega_0 < m\Omega_H.
\ee 

\section{Summary}
\label{sec:sum}

In this article, we have studied the superradiant instability of bound state charged massive scalar fields in Kerr-Sen black holes background. We show analytically how an imaginary part arises in the corresponding scalar's frequency $\omega$ which leads to an amplification factor of the fields. In showing this instability, we adopt the method by Furuhashi and Nambu in Ref. \cite{Furuhashi:2004jk}, where the unstable modes of the bound state charged massive scalar perturbation in the Kerr-Newman black hole background is studied. Hence, since Kerr-Newman and Kerr-Sen black holes appear to have several similarities in their physical properties, it is natural to expect that a charged and massive scalar perturbation is also unstable in the Kerr-Sen black hole background. In addition, we support the findings in literature that a charged and massive scalar perturbation could gain the superradiance effect in the background of GMGHS black hole, i.e. non-rotating limit of Kerr-Sen black hole.

A further study on the (in)stability of the bound state scalar field in the Kerr-Sen or Kerr-Newman black holes can be performed by examining the effective potential in the Shcroedinger-like equation for the radial part of the corresponding wave function. Does the potential really form a ``well'' that could trap the scalar field which escape from the black hole's ergosphere? If it does, the bouncing back and forth of scalar fields with superradiant instability could take place between the ergosphere and some outer points which leads to the ``bomb'' effect of black holes\cite{x2}. In Refs. \cite{Hod:no-bomb,ran-li}, the authors compute the extremum of effective potential $V(r;M,Q,\mu,e,\omega,l)$ in the associated Schrodinger-like radial equation of scalar field in the background of Reissner-Nordstrom (RN) and GMGHS black holes. They found that the arising effective potential cannot provide a suitable well in creating the bomb effect for these black holes. Even though the superradiance occurs for charged scalar perturbation in the background of RN and GMGHS black holes, the bouncing back and forth of scalar fields between ergosphere and some points outside the horizon does not exist for these charged black holes\cite{Hod:no-bomb,ran-li}. It is also interesting to study the Kerr-Sen black hole with scalar hair as well as the scalar clouds related to this black hole \cite{Herdeiro}.

However, Kerr black holes with a massive scalar field perturbation appear to have the ``bomb'' effect. The mass term in the corresponding effective potential of the radial Schrodinger-like equation plays a role like a artificial mirror which is put to yield the ``bomb'' effect \cite{Cardoso:bomb,Hod:analytic}. Therefore, intuitively we guess that the same outcome can be found for rotating and charged black holes when they are impinged by a (charged) massive scalar field. An analytic proof to this manner by studying the effective potential extremum in the equation of motion is addressed in our future work.

\section*{Acknowledgments}

I thank my colleagues at physics dept. UNPAR for their support. I also thank the authors of \cite{Kokkotas:2015uma} for their comments on the misleading statements\footnote{On the state of scalar fields under discussion, in the bound state or not. The instability due to superradiance effect occurs if the scalar fields in a rotating black holes background are in the bound state.} in the previous version of this manuscript.

\appendix

\section{Whittaker Function}\label{app:Witt}

Whittaker function of the first kind $U(r) = M_{\alpha,\beta}(r)$ and the second kind $U(r) = W_{\alpha,\beta}(r)$ are the solutions to the equation \cite{Bailey}
\be
\frac{{d^2 U}}{{dr^2 }} + \left[ { - \frac{1}{4} + \frac{\alpha }{r} - \frac{{1 - 4\beta ^2 }}{{4r^2 }}} \right]U = 0.
\ee
These functions explicitly can be read as
\be
M_{\alpha ,\beta } \left( r \right) = e^{ - r/2} r^{\beta  + 1/2} M\left( {\beta  - \alpha  + 1/2;2\beta  + 1;r} \right),
\ee
and
\be 
W_{\alpha ,\beta } \left( r \right) = e^{ - r/2} r^{\beta  + 1/2} U\left( {\beta  - \alpha  + 1/2;2\beta  + 1;r} \right).
\ee
In equations above we have used the confluent hypergeometric functions $M(a;b;r)$ and $U(a;b;r)$ whose definitions are
\be
M\left( {a;b;r} \right) = _1 F_1 \left( {a;b;r} \right) = \sum\limits_{n = 0}^\infty  {\frac{{a_{\left( n \right)} r^n }}{{b_{\left( n \right)} n!}}} 
\ee
and
\be
U\left( {a;b;r} \right) = \frac{{\Gamma \left( {1 - b} \right)M\left( {a;b;r} \right)}}{{\Gamma \left( {1 + a - b} \right)}} + \frac{{\Gamma \left( {b - 1} \right)M\left( {1 + a - b;2 - b;r} \right)r^{1 - b} }}{{\Gamma \left( a \right)}}.
\ee
Related to our discussion on far region approximation in section \ref{sec:far-region}, the following asymptotic formulas at $r\to \infty$
\be
M_{\alpha ,\beta } \left( r \right)\sim \frac{{\Gamma \left( {2\beta  + 1} \right)r^{ - \alpha } e^{r/2} }}{{\Gamma \left( {\beta  - \alpha  + 1/2} \right)}}\,\,\,\,{\rm and}\,\,\,\,W_{\alpha ,\beta } \left( r \right)\sim r^\alpha  e^{ - r/2} 
\ee
are found to be useful. 

\section{Near and far region solutions matching}
Suppose that the small $r$ approximation of radial solution in far region can be written as
\be
R_{far}  = D_1 \left[ {f_C \left( {\omega ,\mu ,M,e,Q} \right)r^l  + f_D \left( {\omega ,\mu ,M,e,Q} \right)r^{ - l - 1} } \right]
\ee
and the large $r$ approximation of radial solution in near region is
\be
R_{near}  = A\left[ {f_A \left( {\omega ,\mu ,M,e,Q} \right)r^l  + f_B \left( {\omega ,\mu ,M,e,Q} \right)r^{ - l - 1} } \right].
\ee
In matching the two functions above, the following relations hold
\be 
D_1 f_C  = Af_A\,\,;\,\, D_1 f_D  = Af_B \,\,;\,\,\frac{{f_A }}{{f_C }} = \frac{{f_B }}{{f_D }}.
\ee


\end{document}